# Switching From Reactant to Substrate Engineering in the Selective Synthesis of Graphene Nanoribbons

Néstor Merino-Díez,[1,2,3] Jorge Lobo-Checa,[4,5] Pawel Nita,[1,2] Aran Garcia-Lekue,[1,6] Andrea Basagni,[7] Guillaume Vasseur,[1,2] Federica Tiso,[7] Francesco Sedona,[7] Pranab K. Das,[8,9] Jun Fujii,[8] Ivana Vobornik,[8] Mauro Sambi,[7,10] José Ignacio Pascual,[3,6] J. Enrique Ortega,[1,2,11] and Dimas G. de Oteyza [1,2,6,*]

[1] Donostia International Physics Center (DIPC), 20018 San Sebastián, Spain.

[2] Centro de Física de Materiales (CSIC-UPV/EHU) - MPC, 20018 San Sebastián, Spain.

[3] CIC nanoGUNE, Nanoscience Cooperative Research Center, 20018 San Sebastián-Donostia, Spain.

[4] Instituto de Ciencia de Materiales de Aragón (ICMA), CSIC-Universidad de Zaragoza, 50009 Zaragoza, Spain.

[5] Departamento de Física de la Materia Condensada, Universidad de Zaragoza, 50009 Zaragoza, Spain.

[6] Ikerbasque, Basque Foundation for Science, 48011 Bilbao, Spain.

[7] Dipartimento di Scienze Chimiche, Università Degli Studi Di Padova, 35131 Padova, Italy.

[8] Istituto Officina dei Materiali (IOM)-CNR, Laboratorio TASC, 34149 Trieste, Italy

[9] International Centre for Theoretical Physics, 34100 Trieste, Italy

[10] Consorzio INSTM, Unità di Ricerca di Padova, 35131 Padova, Italy

[11] Departamento de Física Aplicada I, Universidad del Pais Vasco, 20018 San Sebastián, Spain.

e-mail: d_g_oteyza@ehu.es

## Abstract

The challenge of synthesizing graphene nanoribbons (GNRs) with atomic precision is currently being pursued along a one-way road, based on the synthesis of adequate molecular precursors that react in predefined ways through self-assembly processes. The synthetic options for GNR generation would multiply by adding a new direction to this readily successful approach, especially if both of them can be combined. We show here how GNR synthesis can be guided by an adequately nanotemplated substrate instead of by the traditionally designed reactants. The structural atomic precision, unachievable to date through top-down methods, is preserved by the self-assembly process. This new strategy´s proof-of-concept compares experiments using 4,4´´-dibromo-para-terphenyl as molecular precursor on flat Au(111) and stepped Au(322) substrates. As opposed to the former, the periodic steps of the latter drive the selective synthesis of 6 atom-wide armchair GNRs, whose electronic properties have been further characterized in detail by scanning tunneling spectroscopy, angle resolved photoemission and density functional theory calculations.

## TOC Graphic

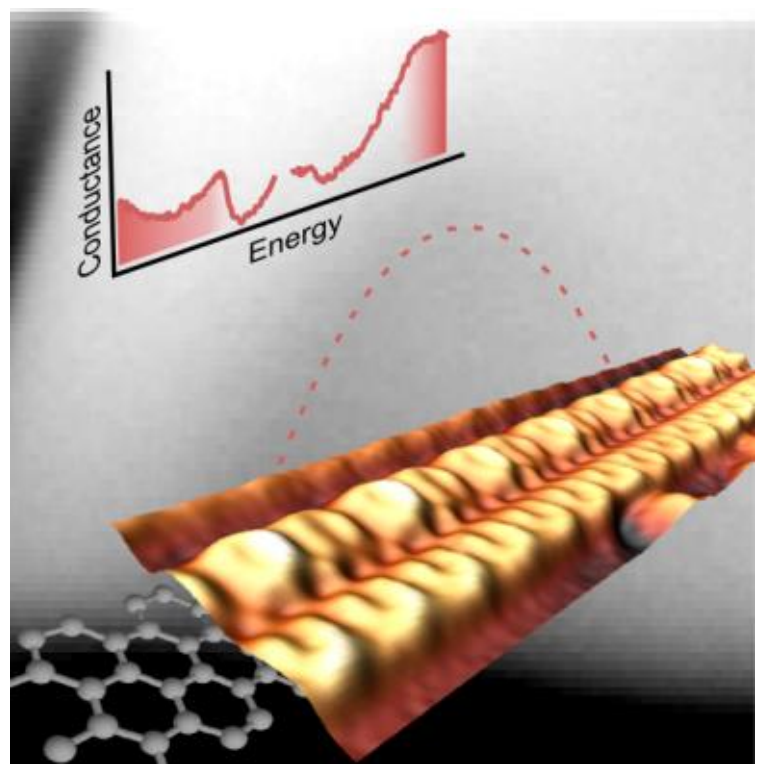

The intensively sought integration of graphene into optoelectronic devices has driven, among other forefront research lines, the thriving quest of graphene nanoribbon (GNR) synthesis.[1–3] Confining graphene into narrow, one-dimensional structures adds a powerful handle to the tunability of its electronic properties.[1,2,4] Small changes in the GNR´s width,[5–10] edge orientation[5,11–14] or termination,[15–17] as well as the controlled addition of heteroatoms,[17–23] can lead to dramatic changes in properties like its band gap (ranging from metallic to wide band gap semiconductors), charge carrier mobility or energy level alignment. However, the experimental synthesis of GNRs with the required atomic precision remains challenging. To date, atomically precise nanoribbons can only be synthesized from the bottom-up, getting appropriately designed precursors to react in pre-defined ways that end in GNR formation.[1–3] As a consequence, great efforts have been placed on the rational design of adequate reactants. But while such efforts have led to impressive results,[5–12,16–23] the number of GNR structures that can be synthesized selectively and with atomic precision is still relatively limited. That pool of GNRs, and consequently their future integration into devices, may develop much faster if new approaches towards selective synthesis become available, especially if the different approaches can be even combined.

An alternative route towards atomically precise GNR structures has been the lateral fusion of GNRs through cyclodehydrogenation,[24–28] although the examples reported to date all suffered from a lack of selectivity. Inspired by previous substrate-guided “on-surface synthesis” examples,[29] this work proves the feasibility of a so far unexplored strategy in GNR growth. Still using a bottom-up approach to guarantee the atomic precision, based on the fusion of neighboring molecular structures, we switch from reactant to substrate engineering in the GNR design. That is, the selectivity in the synthesis process now is triggered and guided by the nanotemplating effect of an adequate substrate.

This is demonstrated using 4,4”-dibromoterphenyl (DBTP) as molecular precursor and two different gold surfaces as substrate, namely Au(111) and Au(322). As the DBTP-decorated surfaces are annealed, the molecules first polymerize into poly-paraphenylene (PPP).[20,27,28,30] At higher temperatures, neighboring PPP chains laterally fuse together and form armchair-oriented graphene nanoribbons (aGNRs).[27,28] Following the same experimental procedures on both surfaces, scanning tunneling microscopy (STM) reveals that a flat Au(111) surface ends up decorated with few remnant PPP wire segments plus a disordered mixture of aGNRs of varying width.[27,28] Instead, on Au(322), which features regularly spaced terraces, apart from the remnant PPP wires, only homogeneous and uniaxially aligned GNRs are formed. These are identified as selectively synthesized aGNR with 6 dimer lines across their width (6-aGNR). Such samples further provide excellent conditions for a detailed characterization of their electronic properties not only by scanning tunneling microscopy and spectroscopy (STM/STS), but also by high resolution angle-resolved photoemission (ARPES), whereby we access key properties of utmost interest for eventual applications like the frontier band´s energy level alignment and effective mass.

DBTP molecules adsorbed on Au surfaces are known to undergo a series of chemical reactions upon annealing (Fig. 1a).[27,28] In our experiments we first sublimate the molecules onto surfaces held at RT. We then anneal the samples to 180 ºC to trigger the polymerization of DBTP into PPP through Ullmann coupling, whereby single Br atoms are left on the surface as byproducts, typically forming rows sandwiched between the PPP chains.[20,27,28] Further increasing the temperature, the Br atoms desorb from the substrate, allowing the PPP chains to approach each other and fuse as the cyclodehydrogenation is activated at temperatures around 380 ºC.[27,28] As a result, wider aGNRs form whose width is determined by the number of participant PPP wires. Quantified by the number of dimer lines across the aGNR (Fig. 1a), the resultant widths thus correspond to multiples of three (3n-aGNR, n being the number of fused PPP). Since aGNRs are classified into three families depending on their number of dimer lines (3p-1, 3p and 3p+1, p being an integer),[4,31,32] all nanoribbons synthesized in these experiments thus correspond to the same 3p family.

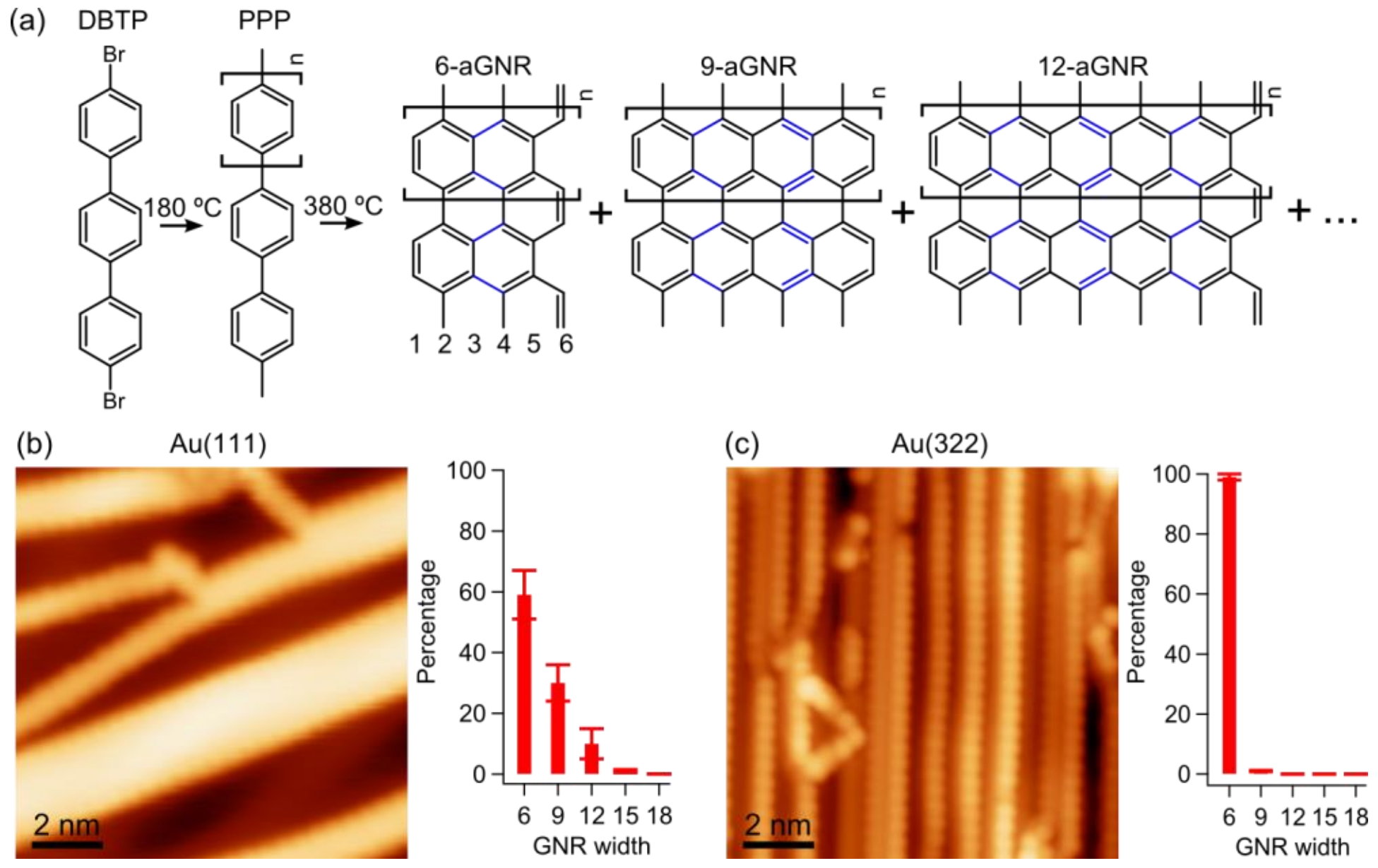

**Figure 1**. Bottom-up synthesis of 3n-aGNR from DBTP. (a) Molecular structure diagram of the reactant and the resulting products after each annealing step. (b) Constant current STM image (10×10 nm$^2$, I = 50 pA, U = -1 V) of the sample on Au(111), decorated with a disordered mixture of differing width aGNRs. The associated histogram displays the percentage of PPP consumed in the formation of GNRs of each different width. (c) Constant current STM image (10×10 nm$^2$, I = 500 pA, U = -1.5V) of the sample on Au(322), evidencing the selective growth of uniaxial 6-aGNR by substrate templating. The histogram shows how all PPP undergoing cyclodehydrogenation (around 50 % in this sample) is used up in the sole formation of 6-aGNRs.

Experiments performed on flat Au(111) evidence that the first polymerization brings about the formation of large islands with long PPP wires oriented along the substrate´s $[10\bar{1}]$ and equivalent directions.[27] However, the second cyclodehydrogenation step at higher temperatures results in disordered, randomly oriented GNRs of varying width, accompanied by some unreacted PPP (Fig. 1b).[27,28] The GNR widths typically range from 6 to 15 dimer lines, decaying in frequency for increasing width. This can be extracted from the histogram in Fig. 1b, which displays the percentage of PPP consumed in the formation of the different GNRs.

A completely different scenario is found on the stepped Au(322) surface. This surface is characterized by uniaxially aligned and regularly spaced terraces, whose steps run along the $[10\bar{1}]$ direction.[33] In addition to the natural templating effect of the steps,[34–37] the favored PPP growth direction coincides with that of the terraces, thus strongly promoting the uniaxial growth of PPP parallel to the substrate steps.[27] The terraces of Au(322) are characterized by an average width of ~12 Å,[33,38] which can fit side-by-side two PPP chains at most. This was initially expected to drive a selective pair-wise fusion of PPP, and the result indeed shows the desired selectivity, displaying 6-aGNRs as the only product (Fig. 1c). It should be noted, however, that the average length of defect free 6-aGNRs is relatively short, namely in the order of 6 nm, typically terminated by 6-aGNR/PPP junctions (including e.g. the shortest possible PPP segments arising from a missing phenyl ring in 6-aGNRs). Nevertheless, this length is sufficient for the GNRs to readily display electronic properties close to those of their infinite analogues.[39] In essence, making use of reactions that can form a variety of different products (Fig. 1a,b), it is the appropriate substrate which imposes synthetic selectivity of 6-aGNR and at the same time their unique azimuthal alignment (Fig. 1c).

However, as displayed in Fig. 2, looking at the sample at different stages of the reaction, the process is found to be more complicated than anticipated. Upon DBTP deposition, the periodic Au(322) terraces (Fig. 2a) act as expected, driving the self-assembly of the pristine molecules into a highly ordered structure with all molecules uniaxially aligned parallel to the steps and with two side-by-side rows of molecules fitting each terrace (Fig. 2b). Five lobes can be clearly distinguished along each DBTP molecule in the STM images, the two outer ones corresponding to the Br atoms and the three inner ones to the three phenyl rings. Upon polymerization, the STM imaging reveals most surprising changes. Polymerized structures with periodic lobes corresponding to the phenyl-units along the PPP are clearly recognized, separated by rows of round objects that correspond to Br atoms (Fig. 2c).[20,27,28,40,41] However, the underlying surface appears completely reconstructed, hosting the alternating rows of PPP and Br on much wider and irregular terraces. This reconstruction is associated to the strong interaction of the halogen atoms with the stepped Au substrate,[42] but will not be discussed further as it is beyond the scope of this work. At first sight, this fact may be expected to hinder the templating effect of the Au(322) surface. However, as bromine desorbs upon further annealing, the Au(322) periodicity is recovered and the substrate templating effect sets in, resulting in the selective production of 6-aGNRs only accompanied by unreacted PPP (Fig. 2d).

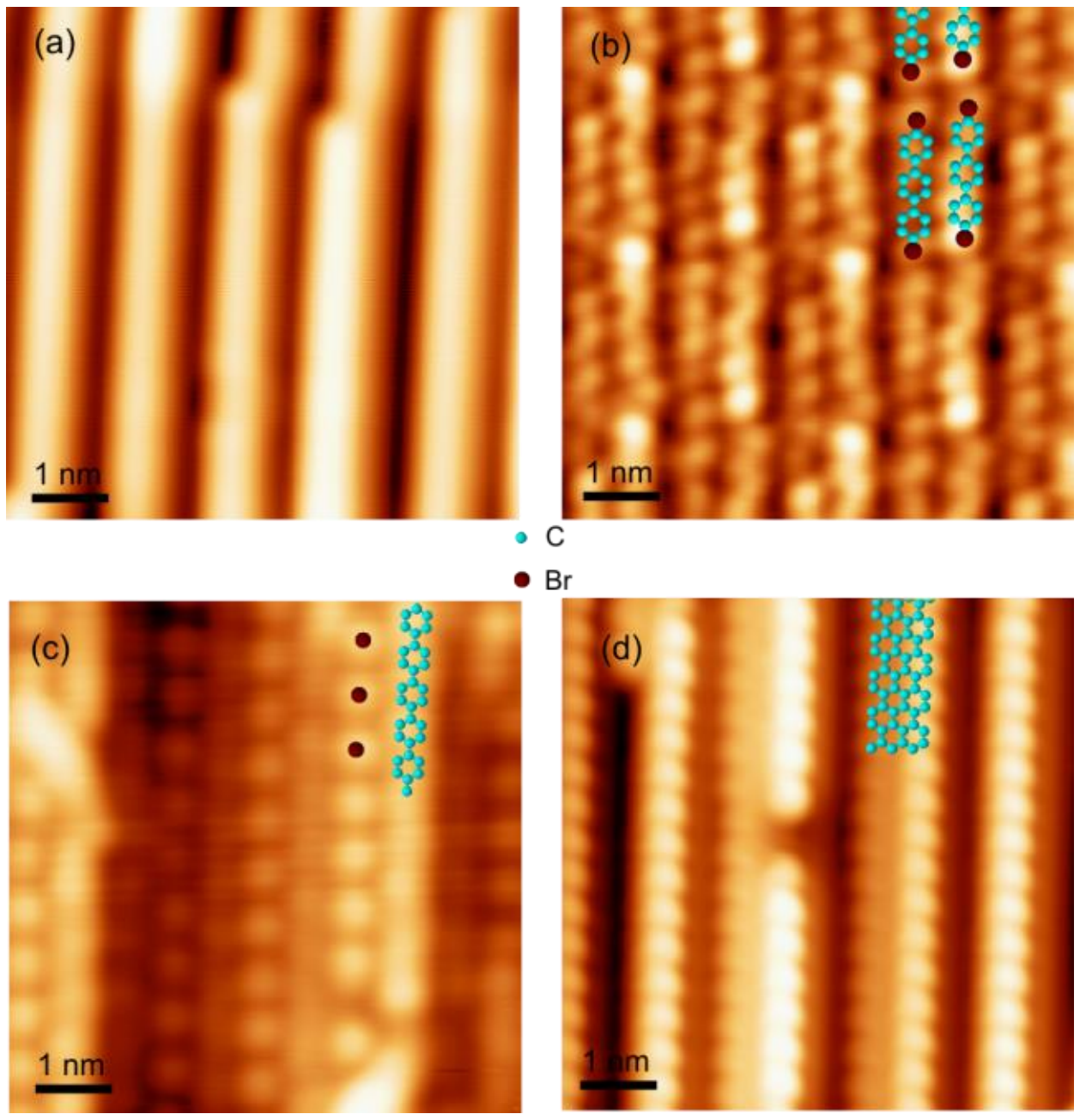

**Figure 2**. Selected 6×6 $nm^2$ constant current STM images at different stages of the growth process: (a) Clean Au(322) substrate (I = 2 nA, U = -5 mV), (b) after DBTP deposition (I = 74 pA, U = 25 mV), (c) after Ullmann polymerization (I = 29 pA, U = 86 mV), and (d) after 6-aGNR formation through cyclodehydrogenation of neighboring PPP chains (I = 516 pA, U = -1.5 V). Molecular structures are overlaid on part of the images as a guide to the eye.

From a characterization point of view, the resulting uniaxially aligned products have the virtue of allowing the use of laterally averaging techniques like angle-resolved photoemission without losing the momentum resolution. Thus, we have used ARPES to monitor the valence band (VB) structure of the vicinal sample at different stages of the GNR synthesis process. Figure 3 displays the electron dispersion along the terraces for the substrate before and after deposition of the reactant, as well as after different annealing treatments. From comparison to the clean Au(322) reference (Fig. 3a), the as-deposited DBTP is observed to produce distinct intensity at an energy of -1.78±0.05 eV (Fig. 3b). This signal is associated to its highest occupied molecular orbital (HOMO), which exhibits a flat band due to the electron´s confinement within the relatively small molecule.[40,43]

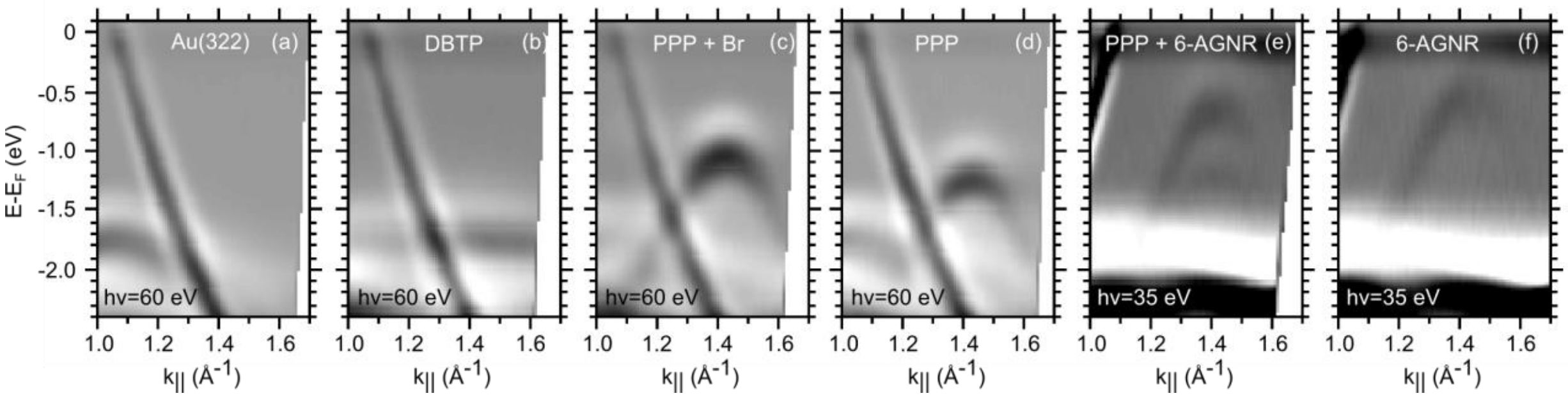

**Figure 3**. Angle-resolved photoemission signal, displaying the dispersion parallel to the step direction, at different stages of the reaction process [integration along $k_\perp$ from 0.2 to 0.4 Å$^{-1}$ and a second derivative processing of the spectral functions have been applied for optimum visualization (see Fig. S1 for the associated raw data, together with the fitted parabolic profiles used to extract the band onset and effective mass)]: (a) Reference spectrum for the bare Au(322) substrate, (b) after DBTP deposition, (c) after Ullmann polymerization into PPP, (d) after Br desorption, (e) after partial fusion of PPP into 6-aGNRs, and (f) after a maximized transformation of PPP into 6-aGNRs. All correspond to different samples heated to increasingly high temperatures, starting from a DBTP covered sample after molecular deposition onto a substrate held at RT. The photon energy used is 60 eV for panels (a-d) and 35 for panels (e-f) to enhance the signal of the existing organic bands. Data on PPP also measured with a photon energy of 35 eV are additionally displayed in Fig. S1.

After Ullmann polymerization, the band structure changes substantially with the appearance of a strongly dispersive band whose topmost states are at E = -1.09±0.05 eV and $k_{||}$ = 1.43 Å$^{-1}$ (Fig. 3c). The dispersive behavior now stems from the electron delocalization along the π-conjugated PPP chain.[40,43] Because the band gap of a conjugated polymer scales approximately with a 1/N relation,[44,45] N being the number of conjugated electrons, the increased conjugation length when going from a precursor with three conjugated rings to a polymer with tens of them is equally responsible for the reduction of the adsorbate´s band gap. As a consequence the frontier

band´s onsets approach the Fermi level, as observed in the ~0.7 eV upshift in energy of the valence band onset with respect to DBTP´s HOMO.

Annealing at higher T (~320 ºC) brings along the desorption of Br [12,27,30,46] and the associated change in work function[12,46] lowers the band onset to E = -1.29±0.03 eV (Fig. 3d) while keeping the effective mass unchanged. The latter may be expected from the absence of chemical changes on the PPP, but it also underlines the limited effect of the intercalated Br chains on the polymers, other than altering the supporting substrate and its associated work function. Annealing to even higher temperature (T > 380ºC) triggers the lateral fusion of PPP chains.[27,28] In the initial stages one can observe the co-existence of the PPP band together with an additional band at higher energy that we associate with the newly formed 6-aGNRs (Fig. 3e). At higher temperature, the PPP band practically disappears at the expense of the new band (Fig. 3f). Other experimental parameters being unchanged, the photoemission signal associated with PPP is proportional to its surface coverage, thus suggesting a transformation of most PPP into 6-aGNRs. It is worth noting that such a high yield has not been achieved in our STM experiments, which always revealed substantial amounts of unreacted PPP (Figs. 1 and 2). Presumably this relates to the different heating rates in the preparation chambers for the ARPES and STM experiments, with a potentially strong impact on the kinetics of this complicated reaction that involves not only the molecular adsorbates, but also important substrate reconstructions.

The VB onset of 6-aGNRs appears at the same momentum as that of PPP, but its energy shifts upward by ~0.6 eV to -0.65±0.08 eV. The common momentum of the VB maxima at 1.43 $Å^{-1}$ is associated to a periodicity of 4.39 Å, in turn related to the adsorbate´s unit cell, the inter-phenyl spacing in PPP and the coincident armchair periodicity of 6-aGNR (Fig. 1). Corresponding to the center of the second Brillouin zone, 1.43 $Å^{-1}$ also coincides with the reciprocal space area where the reactant´s and all product´s photoemission intensity is seen best, since the band´s spectral weight distribution is known to correlate with the Fourier transform of the orbitals, and the HOMO orbitals of the various structures studied are all modulated according to the armchair period.[40,47,48] Thus, since the maximum ARPES intensity along the direction of its long molecular axis is expected around k values of ~1.45 $Å^{-1}$, the reciprocal space region depicted in Figure 2 allows an excellent comparison of the electronic properties of the different adsorbate systems.

PPP, being arguably considered a 3-aGNR, belongs to the same 3p family as 6-aGNR.[28,31,32] Because within each aGNR family the band gap decreases monotonously with increasing width,[28,31,32] the observed upward shift in energy of the VB onset relates again to a decreasing band gap [which brings both valence and conduction band (CB) onsets closer to the Fermi level] when changing from PPP to 6-aGNR. This change in band gap has been measured recently by STS on Au(111).[28] However, because the stepped structure of Au(322) substantially lowers its work function as compared to that of flat Au(111),[13] a notable discrepancy in the energy level alignment is expected. Thus, by STS we have now accessed the band gap and energy level alignment of PPP and 6-aGNRs directly on Au(322) (Fig. S2). Fitting within the error margins with the values obtained from ARPES, the VB onsets measured by STS appear at E = -1.34±0.06 eV and E = -0.79±0.06 eV for PPP and 6-aGNRs, respectively. Meanwhile, the measured band gaps amount to 3.05±0.13 eV and 1.88±0.09 eV (Fig. S2). The offset between the measured VB onsets on Au(111) and Au(322) amounts to 0.25±0.08 eV (0.2±0.06 eV) for PPP and to 0.56±0.1 eV (0.42±0.11 eV) for 6-aGNRs if we compare the STS values on Au(111)[28] with the STS (ARPES) values on Au(322). For PPP the offset equals the 0.25 eV change in work function between the two surfaces,[13] thus closely following an ideal vacuum level pinning scenario.[49] For 6-aGNRs, the slightly larger offset qualitatively still fits the work function change, but is around 0.2 eV larger. This minor increase may arise from additional differences in the interface chemistry, like e.g. an enhanced GNR-substrate hybridization.[49,50] As will be shown later, there are experimental evidences hinting at such enhanced hybridization.

We now focus and deepen the characterization of 6-aGNRs, whose selective synthesis by substrate templating is the key point of this work. Its band gap ($E_g$=1.88±0.09 eV) is in excellent agreement with previous state-of-the-art calculations based on many-body perturbation theory (in particular the GW approximation) and the addition of substrate screening through a classical image charge model.[32] Beyond the energy determination, we have characterized the spatial distribution of valence and conduction band orbitals: experimentally with conductance maps at the corresponding onset energies and theoretically with DFT calculations. These are all summarized in Fig. 4, along with the associated constant current image and STS spectrum.

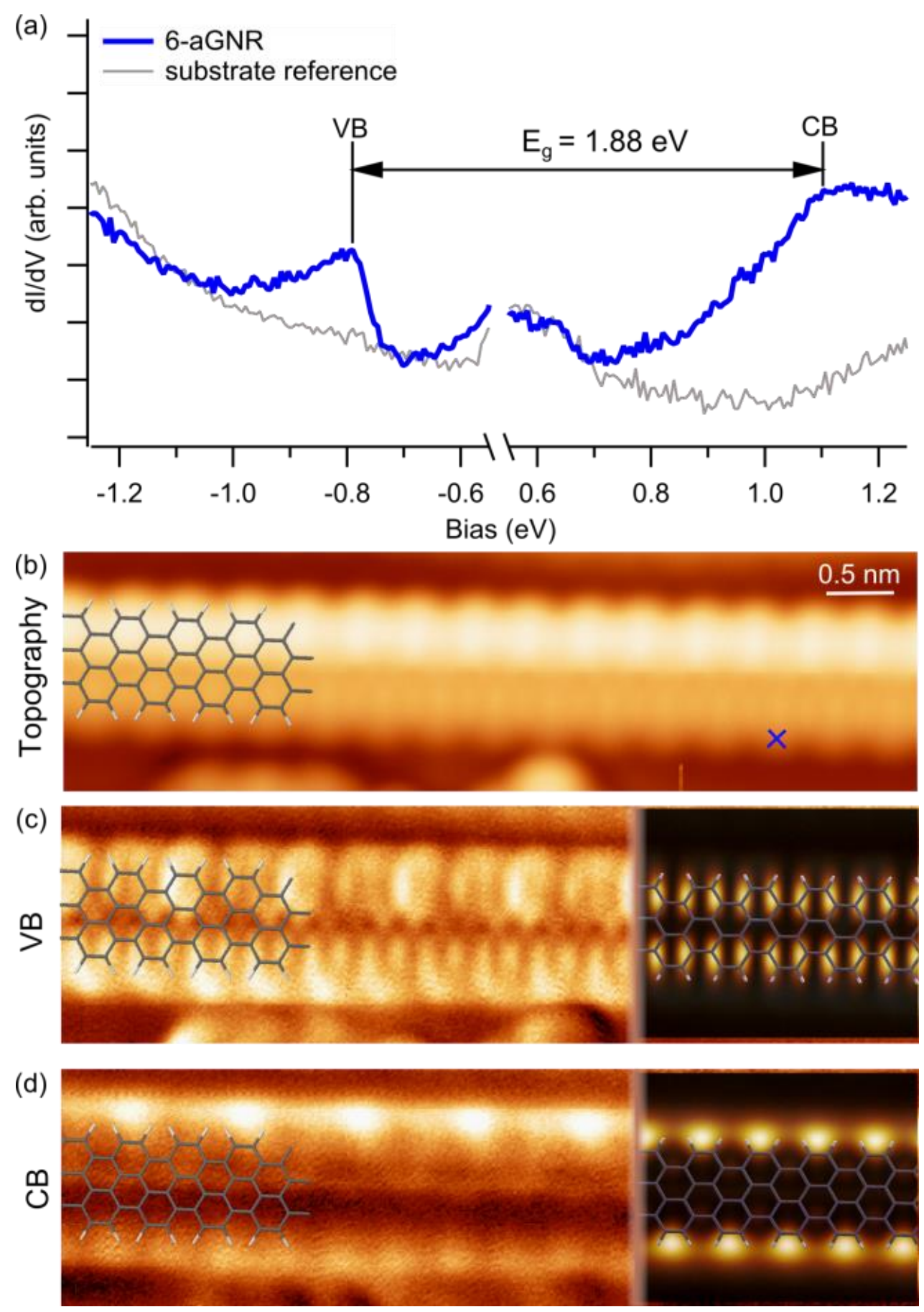

**Figure 4**. Scanning tunneling spectroscopy analysis of the electronic structure of 6-aGNRs. (a) Constant current (I = 430 pA) conductance point spectra on the reference substrate (thin, grey) and on a 6-aGNR (thick, blue), revealing the valence and conduction band onsets. (b) Constant current (topographic) image of a 6-aGNR and the associated conductance maps at the energies of the (c) valence band onset (I = 1.05 nA, U = -0.8 V) and (d) conduction band onset (I = 1.05 nA, U = 1.1 V). The simulated dI/dV images for valence and conduction onsets, evaluated at 4 Å above the C backbone, are super-imposed on the right side of the experimental conductance images for comparison. The molecular structure of the 6-aGNRs is super-imposed on both experimental and simulated dI/dV images as a guide to the eye. A location used for the point spectroscopy on the GNRs (a) is marked by the blue cross in (b).

The simulated conductance images at the energies of valence (Fig. 4c) and conduction band (Fig. 4d) of freestanding 6-aGNRs are evaluated at 4 Å above the molecular plane. Doing so, one accounts for the differently rapid decay towards the vacuum (where the STM tip actually probes the states) of the VB and CB orbitals due to their different wave function symmetries.[9,51] The lacking phase cancellation of the CB orbitals at the GNR sides causes these states to extend further into the vacuum along the ribbon´s edges (Fig. S3). In contrast, the oscillating phase along both the transverse and longitudinal GNR directions of the VB states´ wave functions causes a faster but more homogeneous decay of the LDOS toward the vacuum (Fig. S3).[9,51] As displayed in Figs. 4c,d, taking these effects into account by simulating the conductance maps at 4 Å above the carbon backbone, a good agreement is obtained with the experimental measurements. In addition to complementary constant height dI/dV spectra displaying no increased conductance anywhere around $E_F$ (Fig. S4), such good agreement allows the unambiguous assignment of the observed onsets in Fig. 4a to VB and CB, since the nodal structure and wave function symmetry of the VB-1 and CB+1 are completely different. That is, due to the arguments described above, the VB-1 would be observed strongest along the GNR sides, while the CB+1 would display two nodal planes along the ribbon axis (Fig. S3). The calculations, however, do not reproduce the additional superstructure with twice the armchair unit cell period that is clearly resolved in the experimental images. The superstructure is particularly visible along one of the sides of the GNR, and the reason for it is found in the underlying substrate, not included in our calculations.

Armchair graphene nanoribbons aligned along the compact $[10\bar{1}]$ direction are commensurate every second unit cell.[27,28] On flat Au(111), the molecule-substrate interaction is so weak that STM and STS measurements show no signature of such commensuration in the nanoribbon´s signal.[27,28] However, a different scenario appears as GNRs interact with under-coordinated (and thus more reactive) Au atoms like those at the step edges. Under these circumstances the interactions of Au with GNR and the associated hybridization is stronger, translating into an evident fingerprint of the commensuration in the ribbon´s electronic density of states. The fact that this commensuration fingerprint is more visible along one of the GNR´s sides (top GNR side in Fig. 4) relates to the position of the step edges. Careful image analysis reveals that the GNRs are tilted across two neighboring terraces with the substrate step off-center, that is, closer to one of the two GNR sides (Fig. S5). As a consequence, this particular side will hybridize more strongly with the substrate and thus show a more pronounced imprint of it in the imaging of the GNR orbitals.

Another parameter of utmost relevance in graphene nanoribbons is their effective mass, as it is inversely proportional to the charge carrier mobility along the ribbon.[52] There have been different reports proposing analytical relations between the effective mass and other GNR parameters.[53,54] For example Raza et al. related it to the aGNR width (W, distance in nanometers between C atoms on either edge, calculated based on a C-C bond length of 1.44 Å) through $m^* = 0.091 / W$ for the 3p family, $m^* = 0.160 / W$ for the 3p+1 family and $m^* = 0.005 / W$ for the 3p-1 family.[53] In turn, Arora et al. related it to the aGNR´s bandgap $E_g$ through $m^* = E_g \text{ (eV)} / 11.37 \text{ eV}$.[54] Reported experimental values for the effective mass, measured on atomically precise graphene nanoribbons, are extremely scarce. Apart from chiral graphene nanoribbons, which display a very different band dispersion behavior,[13] the reported values for armchair graphene nanoribbons include those of 6-aGNR characterized in this work, 7-aGNR,[34,35,51,55] the pioneering and to date best studied aGNR synthesized with atomic precision, as well as 9-aGNRs.[9,55] For 7-aGNRs the reported values are scattered in a wide range, depending on the research groups and on the characterization techniques [ARPES, Fourier-transformed STS (FT-STS)]. However, more consistent values from different research groups and experimental techniques are reported for the latter. Table 1 summarizes the effective mass values of 6-aGNRs, 7-aGNRs and 9-aGNRs reported from experiments and those estimated by the above described relations of Raza and coworkers,[53] as well as of Arora and coworkers.[54] For the latter, the bandgap input has been taken both from their calculations (available only for 7-aGNR and 9-aGNR) and from the reported experimental values.[9,51] It can be observed that the estimated effective masses fit within the wide range of experimentally reported values for 7-aGNRs and are in very good agreement with those of 9-aGNRs. Remarkably, an excellent agreement is found as well between the estimations and our experimental measurement on 6-aGNR. Thus, beyond the readily widely accepted band gap predictions for aGNRs, this work supports the reliability of the predictive scaling laws for the effective masses of aGNRs, a similarly important parameter for the ultimate performance of GNR-based devices.

**Table 1**. Effective mass values of aGNRs measured experimentally and estimated theoretically following the relations proposed by Arora et al.,[54] as well as Raza et al.[53]

| | Arora et al. | | Raza et al. | Experiment |
|---|---|---|---|---|
| width (dimer lines) | From calc. $E_g$ | From exp. $E_g$ | | |
| 6 [a] | --- | 0.17±0.01 | 0.15 | 0.15±0.02 (ARPES) |
| 7 | 0.154 | 0.21±0.01[b] | 0.21 | 0.21,[34] 0.22,[55] 1.07,[35] (ARPES)<br>0.41±0.08 [51] (FT-STS) |
| 9 | 0.074 | 0.12 [c] | 0.09 | 0.09±0.02,[9] 0.11,[55] (ARPES)<br>0.12±0.03 [9] (FT-STS) |

a) Experimental values of $E_g$ and m* taken from this work. b) Experimental value of $E_g$ taken from ref.[51] c) Experimental value of $E_g$ taken from ref.[9]

Altogether, we prove a new strategy towards the selective synthesis of GNRs, namely the use of substrate templating. Combining a stepped Au(322) surface and DBTP precursors we report the first selective synthesis of 6-aGNRs.  Furthermore, the uniaxial alignment imposed to the products by the substrate has allowed us characterizing the electronic properties of 6-aGNRs by angle resolved photoemission, in addition to density functional theory calculations and scanning tunneling microscopy/spectroscopy. Thereby, not only the bandgap, but also another important figure of merit in GNRs has been accessed, as is the valence band´s effective mass.

**Methods**

The Au(111) and Au(322) surfaces were cleaned similarly, by standard $Ar^+$ sputtering and annealing cycles. The molecules were then deposited on the surfaces by means of a home-built Knudsen cell heated to ~115 ºC during sublimation. During the deposition the surfaces were held at room temperature, annealed thereafter to subsequently trigger the different reaction steps. The STM characterization was performed at 4.3 K in a commercial Scienta-Omicron LT-STM/AFM. For spectroscopic point spectra and conductance maps, the dI/dV signal was measured by a lock-in amplifier, modulating the bias with 15 mV at 731 Hz. The STM images were processed with the freeware WSxM.[56] The statistics reported in the histograms of Fig. 1 where analyzed as follows: The total length of GNRs of each width was added from multiple large scale images. For each width the total length is multiplied by the number of PPP chains involved in their formation (2 for 6-aGNRs, 3 for 9-aGNRs, etc). After normalization to the total amount of GNRs with 6 or more dimer lines, we obtain the percentage of PPP that undergoes cyclodehydrogenation during the formation of GNRs of each different width.

The ARPES experiments were performed in-situ at the APE-LE beamline of the Elettra Sincrotrone-Trieste, using linearly polarized light. For the ARPES acquisition on 6-aGNRs, the photon energy was tuned from 60 eV to 35 eV because it yielded a better signal-to-noise ratio.

While this does not affect the 1D dispersion of the nanoribbons, the 3D-dispersive substrate-related signal changes significantly. Nevertheless, the observed band is unquestionably associated to 6-AGNRs, as can be inferred comparing the emission from the PPP, mixed PPP/6-AGNR and 6-AGNR samples at the same photon energy (Fig. S1, including second derivative images for a better visualization of the bands). The temperature for ARPES acquisition was 90 K and the overall energy and angle resolution was better than 30 meV and 0.1°.

The optimized geometry and electronic structure of a free-standing 6-aGNR, composed of 6 carbon dimer lines passivated with hydrogen at the edges, were calculated using density functional theory as implemented in the SIESTA code.[57] The 6-aGNR was relaxed until forces on all atoms were $< 0.01$ eV/Å, and the dispersion interactions were taken into account by the non-local optB88-vdW functional.[58] The basis set consisted of double-zeta plus polarization (DZP) orbitals for both species, with an Energy Shift parameter of 50 meV. A 18x1x1 Monkhorst-Pack mesh was used for the k-point sampling of the three-dimensional Brillouin zone, where the 9 k-points are taken along the direction of the ribbon. A cutoff of 300 Ry was used for the real-space grid integrations. The simulated STM images were obtained using the STM utility in SIESTA, which allows calculating the actual space charge density at experimentally realistic distances above the graphene nanoribbon.

**Acknowledgements**

The project leading to this publication has received funding from the European Research Council (ERC) under the European Union's Horizon 2020 research and innovation programme (grant agreement No 635919), from the Spanish Ministry of Economy, Industry and Competitiveness (MINECO, Grant No. MAT2016-78293-C6), from the Basque Government (Grant No. IT-621-13), from the regional Government of Aragon (RASMIA project) and from the University of Padova (Grant CPDA154322, Project AMNES). The work has been partly performed in the framework of the nanoscience foundry and fine analysis (NFFA-MIUR Italy Progetti Internazionali) facility.

**Supplementary Information.** Raw ARPES data of PPP, partially fused PPP and quasi-entirely fused 6-aGNRs. Comparison of dI/dV spectra of PPP and 6-aGNR. Calculated wavefunctions of valence and conduction band in free-standing 6-aGNRs and associated height-dependent STM image simulations. Constant height dI/dV spectrum of 6-aGNRs. STM images and profiles evidencing the tilt of 6-aGNRs across substrate steps.

# Supplementary Information:

# Switching From Reactant to Substrate Engineering in the Selective Synthesis of Graphene Nanoribbons

Néstor Merino-Díez,[1,2,3] Jorge Lobo-Checa,[4,5] Pawel Nita,[1,2] Aran García-Lekue,[1,6] Andrea Basagni,[7] Guillaume Vasseur,[1,2] Federica Tiso,[7] Francesco Sedona,[7] Pranab K. Das,[8,9] Jun Fujii,[8] Ivana Vobornik,[8] Mauro Sambi,[7,10] José Ignacio Pascual,[3,6] J. Enrique Ortega,[1,2,11] and Dimas G. de Oteyza [1,2,6,*]

[1] Donostia International Physics Center (DIPC), 20018 San Sebastián, Spain.

[2] Centro de Física de Materiales (CSIC-UPV/EHU) - MPC, 20018 San Sebastián, Spain.

[3] CIC nanoGUNE, Nanoscience Cooperative Research Center, 20018 San Sebastián-Donostia, Spain.

[4] Instituto de Ciencia de Materiales de Aragón (ICMA), CSIC-Universidad de Zaragoza, 50009 Zaragoza, Spain.

[5] Departamento de Física de la Materia Condensada, Universidad de Zaragoza, 50009 Zaragoza, Spain.

[6] Ikerbasque, Basque Foundation for Science, 48011 Bilbao, Spain.

[7] Dipartimento di Scienze Chimiche, Università Degli Studi Di Padova, 35131 Padova, Italy.

[8] Istituto Officina dei Materiali (IOM)-CNR, Laboratorio TASC, 34149 Trieste, Italy.

[9] International Centre for Theoretical Physics, 34100 Trieste, Italy.

[10] Consorzio INSTM, Unità di Ricerca di Padova, 35131 Padova, Italy.

[11] Departamento de Física Aplicada I, Universidad del Pais Vasco, 20018 San Sebastián, Spain.

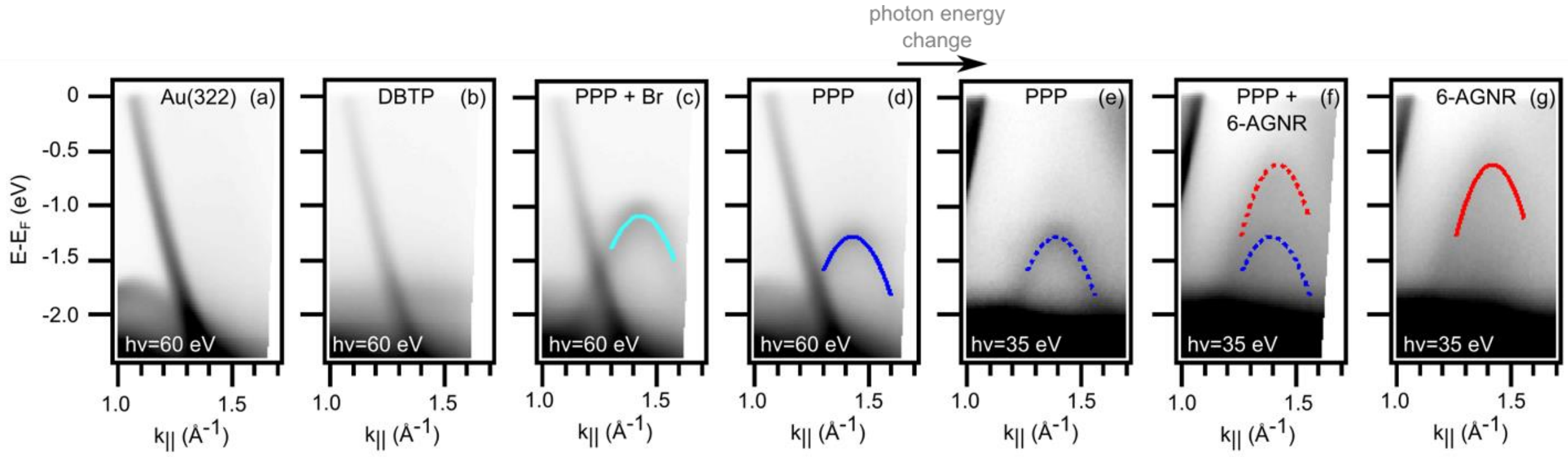

**Figure S1**. Angle-resolved photoemission intensity associated to the processed data in Fig. 3 of the main text, displaying the dispersion parallel to the step direction at different stages of the reaction process (integration along $k_{\perp}$ from 0.2 to 0.4 Å$^{-1}$ has been applied for optimum visualization): (a) Reference spectrum for the bare Au(322) substrate, (b) after DBTP deposition, (c) after Ullmann polymerization into PPP, (d) after Br desorption, (e) after partial fusion of PPP into 6-aGNRs, and (f) after a maximized transformation of PPP into 6-aGNRs. All correspond to different samples heated to increasingly high temperatures [except (d) and (e), which correspond to the same sample measured at different photon energies], starting from a DBTP covered sample after molecular deposition onto a substrate held at RT. The parabolic profiles fitted to the dispersive bands used to extract band onset and effective mass are overlaid on the images. The dotted profiles in panels (e) and (f) are not fitted, but included as a superposition of the profiles in (d) and (g).

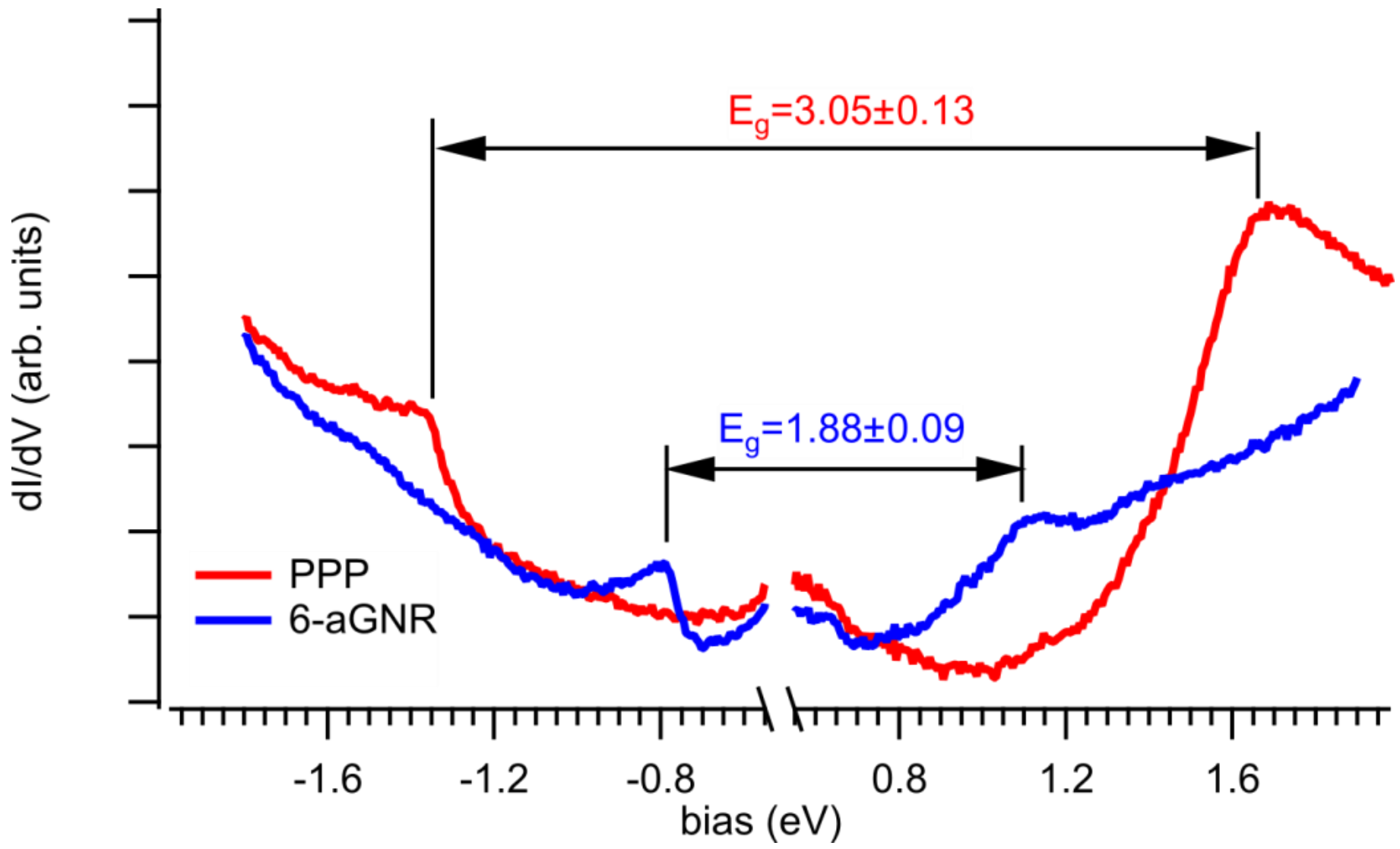

**Figure S2**. Constant current (I = 430 pA) dI/dV spectra of PPP and 6-aGNRs on Au(322), revealing their correspondingly different band gaps and band onset energies.

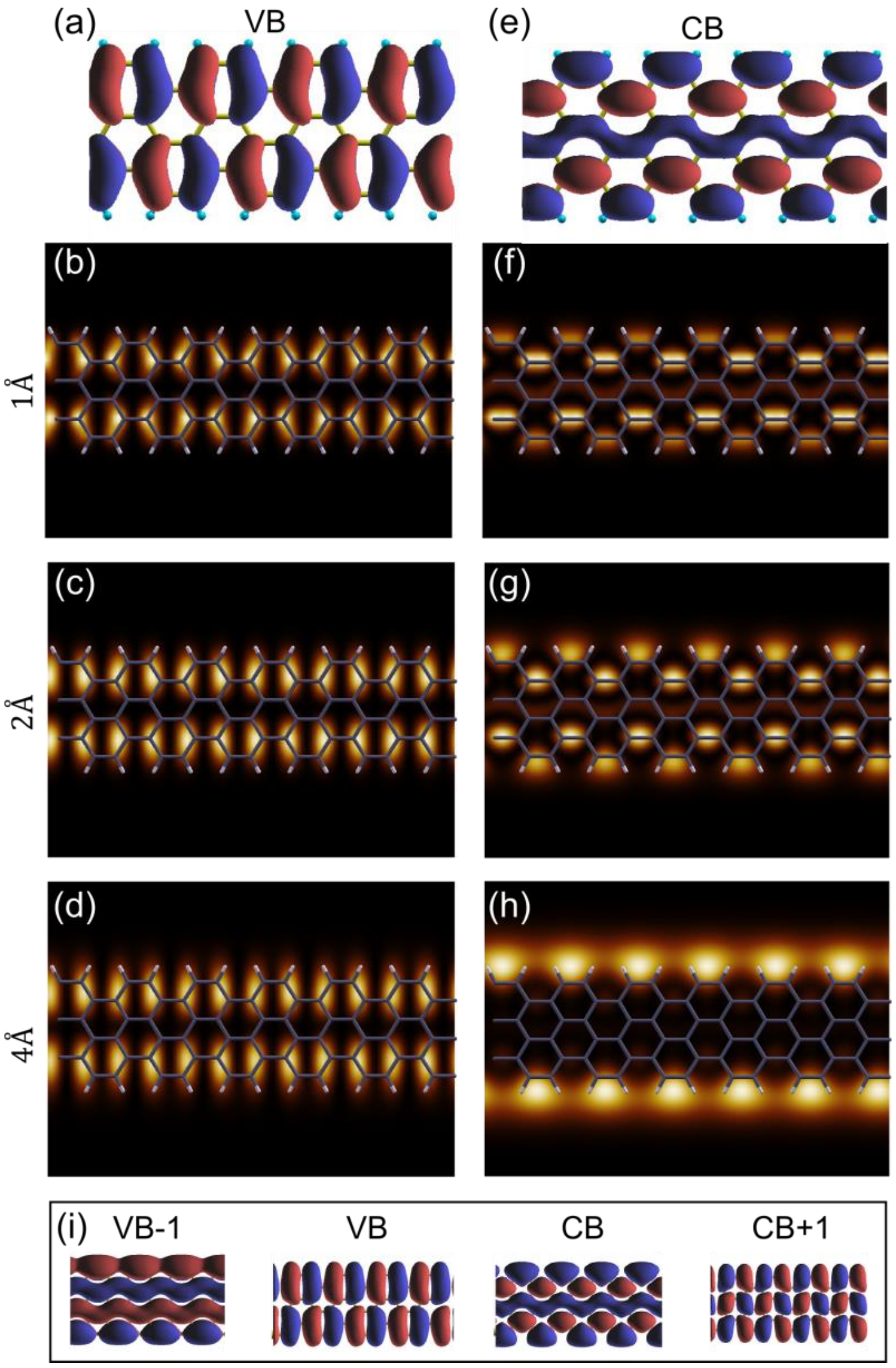

**Figure S3**. DFT calculations displaying the differently fast orbital decay towards vacuum for VB and CB. The wave function of the VB is displayed in (a), and the simulated STM images integrating 100 mV around its onset at distances of 1 Å, 2 Å and 4 Å above the carbon backbone are displayed in (b), (c) and (d), respectively. The wave function and equivalently simulated STM images for the CB are displayed in (e), (f), (g) and (h). The contrast in simulated STM images is saturated to their maximum range in each of the cases. The wave functions for VB-1, and CB+1 are added to those of the VB and CB in (i) for comparison of their symmetries, from which the VB-1 is expected to display a similar behavior to the CB, while the CB is expected to behave rather like the VB in terms of their height dependence or the orbital decay toward vacuum.

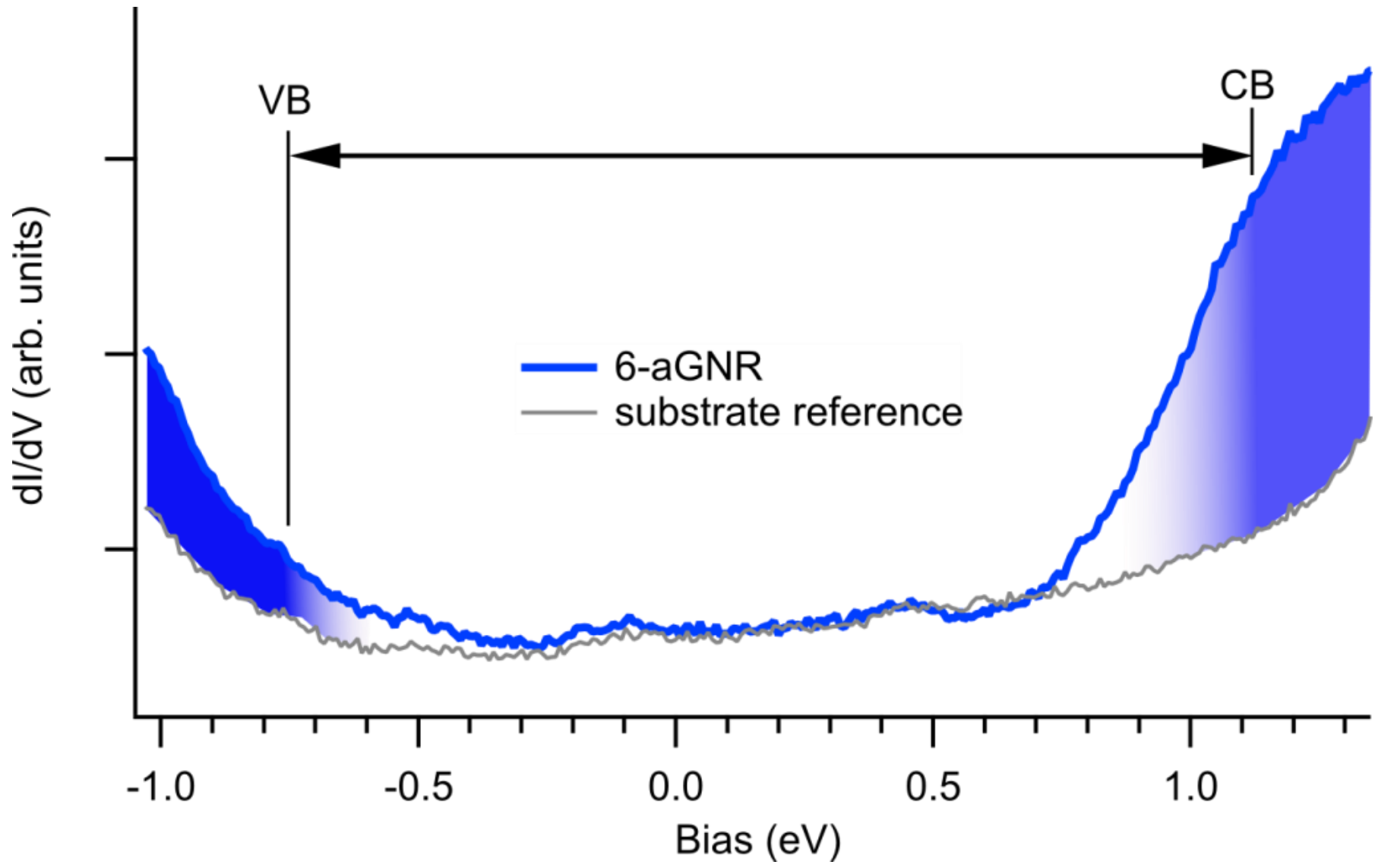

**Figure S4**. Representative constant height dI/dV spectrum of 6-aGNRs. Although with much worse defined band onsets, an increased conductance with respect to the substrate reference (colored in blue) is observed starting from energies similar to those of the better defined onsets obtained from constant current spectra (displayed in Fig. 4 and Fig. S2 and marked with appropriately labelled straight lines). At the same time it confirms the absence of states in the energy range not probed by constant current spectra (further confirmed with the assignment from the dI/dV mapping and the simulations in Fig. S3).

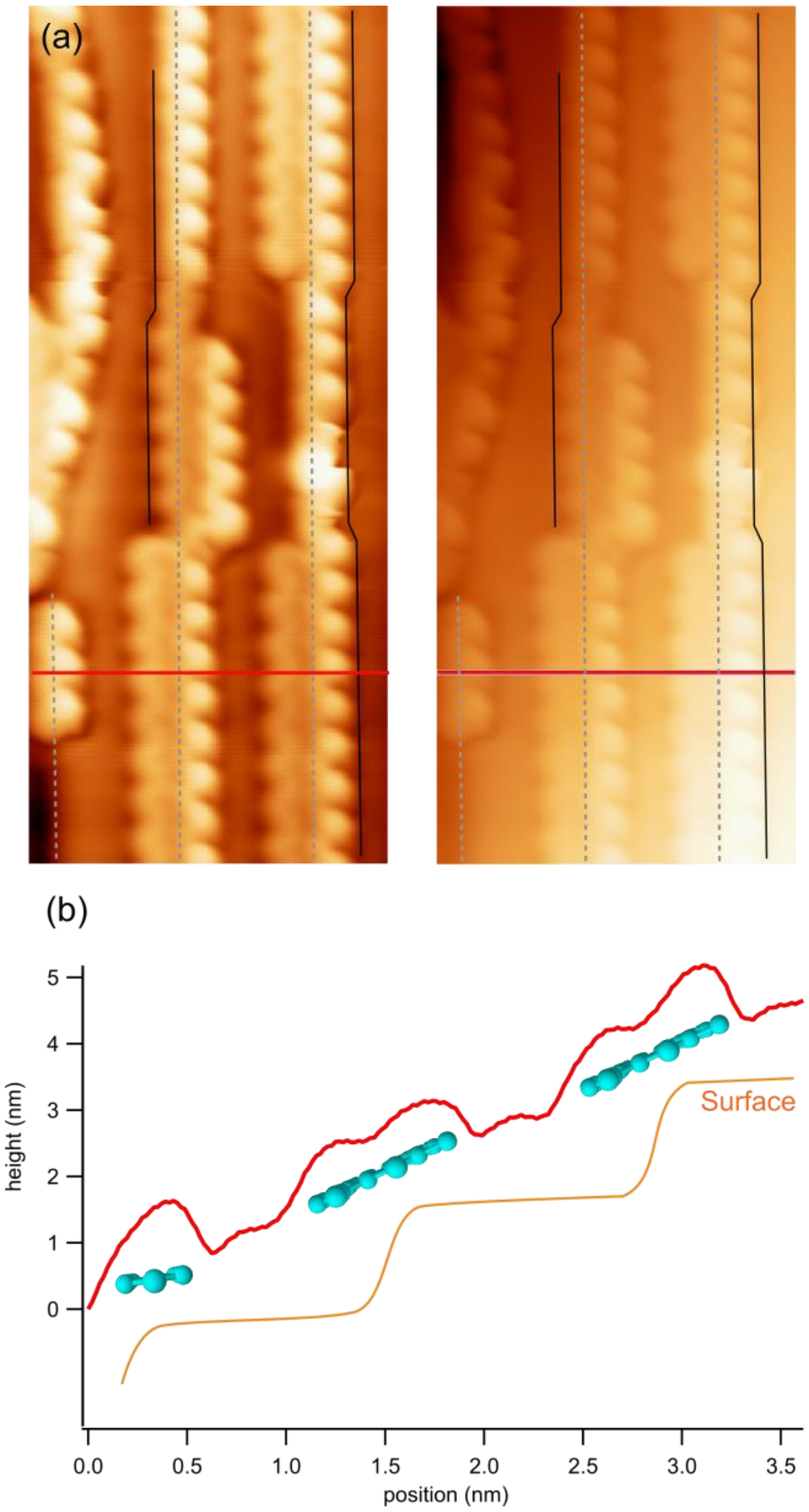

**Figure S5**. Tilted structure of 6-aGNRs. (a) Constant current STM image (with and without a subtracted plane to emphasize the molecular structures or the topography, respectively) marking the step positions (dashed grey lines). Their position with respect to poly-paraphenylene chains are extracted from their imaging on incompletely covered steps (e.g. left step, lower area). Position of adsorbate-decorated steps are in turn marked according to areas covered with PPP and its known alignment relative to the steps. The 6-aGNR edges shift (as compared to the alignment of the connected PPP segments, marked with solid black lines) towards the upper (lower) terrace when they fuse with a second PPP wire from the lower (upper) terrace. The result is a tilted 6-aGNR across the steps of Au(322). (b) A profile along the marked red line in (a), together with an approximate schematic representation of the substrate, PPP and 6-aGNR structures positions. Constant current imaging parameters: I = 102 pA, U = -0.92 V.